\documentstyle[12pt]{article}

\newcommand{\Winf}{$W_{1+\infty}$ }
\newcommand{\be}{\begin{equation}}
\newcommand{\ee}{\end{equation}}
\newcommand{\ba}{\begin{eqnarray}}
\newcommand{\ea}{\end{eqnarray}}
\newcommand{\refs}[1]{(\ref{#1})}
\newcommand {\sfrac}[2]{{\textstyle \frac{#1}{#2}}}
\newcommand {\n}{\nonumber \\}
\newcommand {\ch}{{\rm ch}}

\begin{document}

%%%%%%%%%%%%%%%%%%%%%%%%%%%%%%%%%%%%%%%%
%                                      %
%   Title page                         %
%                                      %
%%%%%%%%%%%%%%%%%%%%%%%%%%%%%%%%%%%%%%%%
\begin{flushright}
January 1994\hfill
YITP/K-1054\\
UT-669\\
SULDP-1994-1\\
hep-th/9402001
\end{flushright}

\vskip 15mm
\begin{center}
{\large\bf
Determinant Formulae\\
\vskip 1mm
of Quasi-Finite Representation of \Winf Algebra\\
\vskip 1mm
at Lower Levels}

\vspace{15mm}

H.~Awata,\hskip 10mm M.~Fukuma\\
Yukawa Institute for Theoretical Physics\\
Kyoto University, Kyoto 606, Japan

\vskip 5mm
Y.~Matsuo\\
Department of Physics\\
University of Tokyo, Tokyo 113, Japan

\vskip 5mm
S.~Odake\\
Department of Physics, Faculty of Liberal Arts\\
Shinshu University, Matsumoto 390, Japan
\end{center}

\vspace{15mm}

\begin{abstract}

We calculate the
Kac determinant for the quasi-finite representation of
\Winf algebra up to level 8.
It vanishes only when the
central charge is integer.
We give an algebraic construction of null states
and propose the character formulae.
The character of the Verma module
is related to free fields in three dimensions
which has rather exotic modular properties.

\end{abstract}

\pagestyle{empty}
\newpage
\pagestyle{plain}
\pagenumbering{arabic}
%%%%%%%%%%%%%%%%%%%%%%%%%%%%%%%%%%%%%%%%
%                                      %
%  1. Introduction                     %
%                                      %
%%%%%%%%%%%%%%%%%%%%%%%%%%%%%%%%%%%%%%%%
\section{Introduction}

In the study of loop algebras,
the Kac determinant \cite{rK} played an essential role for
understanding the detailed structure of the Hilbert space.
In fact, historically we could not have constructed the minimal
models if we had no knowledge about it.
On the other hand, the Kac determinant of the \Winf algebra
has not been discussed so far.
It was mainly due to the fact that infinitely many states possibly appear
at each energy level, reflecting the infinite number of generators.
% $W^1$, $W^2$, $W^3$, $W^4\ldots$.
Although we have explicit representations by free fields \cite{rPRS,rWinf}
and we can even construct its full character formula
in a special case of $C=1$ \cite{rO,rAFOQ},
it remained a longstanding problem
to classify all the possible representations.

Recently,
Kac and Radul \cite{rKR} overcame this difficulty of infiniteness,
explicitly showing the prescription for getting
finite number of non-vanishing states at each energy level.
The representation thus obtained is called quasi-finite.

Now that we have finite number of states at each energy level,
the natural step which should follow
is the computation of the Kac determinant and the character.
In this letter, we report our preliminary results on this subject.
Our computation was carried out by using {\sl Mathematica} package.
We give the Kac determinant up to level 8, where 160 relevant states exist.
We find that additional null states appear
only when the central charge is integer,
the case in which the representation can be realized by
free fermions or bosonic ghosts \cite{rM}.
After explaining the structure of null states,
we then propose the character formulae
which are consistent with the Kac determinant.
Furthermore, we find that the character of the Verma module is related to
the character of {\it three}--dimensional free field
first given by Cardy \cite{rC}.
The modular property of \Winf algebra thus becomes
rather exotic due to this nature.

%%%%%%%%%%%%%%%%%%%%%%%%%%%%%%%%%%%%%%%%
%                                      %
%  2. Definitions of \Winf algebra     %
%                                      %
%%%%%%%%%%%%%%%%%%%%%%%%%%%%%%%%%%%%%%%%
\section{Definitions of \Winf algebra}

Although there are some overlaps
with our previous paper \cite{rM},
we recapitulate some definitions of \Winf algebra and give
a brief review of \cite{rKR}
to make this paper self-contained to some extent.

The classical algebra
% $w_{1+\infty}$ algebra
is generated by
the polynomials of $z$ and $D\equiv z\frac{\partial}{\partial z}$.
One of the typical generator may be written as $z^r f(D)$ where
the function $f(w)$ is any regular function at $w=0$.
% The $w_{1+\infty}$ may be written as,
Their commutation relations are
\be
\left[ z^r f(D), z^s g(D) \right]=
z^{r+s}\left( f(D+s)g(D)-f(D) g(D+r)\right).
\label{eCW}
\ee
We call the quantum version of this algebra the \Winf algebra.
For each classical generator $z^r f(D)$, we denote
the corresponding quantum generator by $W(z^r f(D))$.
The quantum algebra is different from the classical one
\refs{eCW} only in the additional central term,
\ba
  \left[ W(z^r f(D)), W(z^s g(D))\right]
  &\!\!=\!\!&
  W(z^{r+s}f(D+s)g(D))-W(z^{r+s}f(D)g(D+r)) \n
  &&+C \Psi(z^r f(D), z^s g(D)),\label{eQW}
%
%&&\left[ W(z^r f(D)), W(z^s g(D))\right]
%= W(z^{r+s}f(D+s)g(D))\nonumber\\
%&&-W(z^{r+s}f(D)g(D+r))
%+ C \Psi(z^r f(D), z^s g(D)),\label{eQW}
\ea
where two-cocycle $\Psi$ is defined by
\ba
&&\Psi(z^r f(D), z^s g(D))=-\Psi(z^s g(D),z^r f(D))\nonumber\\
&&=\left\{
\begin{array}{ll}
        \sum_{1\leq j\leq r}f(-j)g(r-j) &
        \mbox{if  $r=-s>0$}  \\
        0 & \mbox{if  $r+s\neq 0$ or $r=s=0$}.
\end{array}
\right.
\label{eC}
\ea
These commutation relations can be written in a compact form:
\be
  \left[ W(z^r e^{xD}), W(z^s e^{yD})\right]
  =(e^{xs}-e^{yr})W(z^{r+s}e^{(x+y)D})
  +C\frac{e^{xs}-e^{yr}}{1-e^{x+y}}\delta_{r+s,0}.
  \label{comm}
\ee
The basis given in \cite{rPRS}, $V^i_r=W^{i+2}_r$, are expressed as
$W^{k+1}_r$$=$$W(z^rf^k_r(D))$ ($k\geq 0$),
\be
  f^k_r(D)={2k \choose k}^{-1}
  \sum_{j=0}^k(-1)^j{k \choose j}^2
  \lbrack -D-r-1\rbrack_{k-j} \lbrack D\rbrack_j,
\ee
where $\lbrack x\rbrack_m=\prod_{j=0}^{m-1}(x-j)$ and
${x \choose m}=\lbrack x\rbrack_m/m!$.
The generators of the $\widehat{u(1)}$ and Virasoro\footnote{
$W^2_r=L_r-\frac{1}{2}(r+1)J_r$.}
subalgebras are written as
$J_r=W(z^r)$, $L_r=-W(z^r D)$.
%\label{eVir}
In particular, $L_0$ is used to count the energy level,
$
\left[ L_0, W(z^r f(D))\right]=-r W(z^r f(D)).
%\label{eLW}
$
One may immediately recognize that there are infinitely many
generators at each energy level.

A representation
of \Winf algebra is specified by its highest weight conditions,
\be
\begin{array}{lcl}
        W(z^r D^k)|\lambda\rangle=0, & \qquad &
        r\geq 1, k\geq 0,
        \\
        W(D^k)|\lambda\rangle = \Delta_k |\lambda\rangle ,  &
        \qquad & k\geq 0.
\end{array}
\label{eHWC}
\ee
Unlike other two dimensional algebras, there are
infinitely many parameters
$\Delta_k$ to specify the representation.
In the quasi-finite representation, however, the number
of relevant parameters is reduced to finite.
It will be convenient to introduce,
\begin{equation}
        \Delta(x)\equiv-\sum_{k=0}^\infty {x^k\over k!}\Delta_k.
        \label{eDx}
\end{equation}
This is the eigenvalue of the operator, $W(-e^{xD})$.

%%%%%%%%%%%%%%%%%%%%%%%%%%%%%%%%%%%%%%%%
%                                      %
%  3. Quasi-Finite Representation and  %
%     Character of Verma Module        %
%                                      %
%%%%%%%%%%%%%%%%%%%%%%%%%%%%%%%%%%%%%%%%
\section{Quasi-Finite Representation and Character
of Verma Module}
Kac and Radul studied the structure of the conditions
to get finite number of non-vanishing states at each energy level.
Their result may be summarized as follows.
\begin{enumerate}
        \item   For each level $r$ , generators which annihilate
        the highest weight state should take the form,
        $W(z^{-r} b_r(D)g(D))$ where $b_r(D)$ is a
        monic, finite degree polynomial
        of operator $D$.

        \item   The polynomial $b_r(D)$ with $r>1$ is related to
        level 1 polynomial $b(D)\equiv b_1(D)$ as
        \begin{itemize}
        	\item  $b_r(D)$ can be divided by lcm($b(D)$, $b(D-1)$,
        	$\cdots$ $b(D-r+1)$).

        	\item  $b(D)b(D-1)\cdots b(D-r+1)$ can be
        	divided by $b_r(D)$.
        \end{itemize}
        $b_r$ can be
        uniquely determined as $b_r(D)=\prod_{s=0}^{r-1}b(D-s)$
        only when the differences of any two distinct
        roots of $b(w)=0$ are not integer.

        \item The function $\Delta(x)$
        satisfies a differential equation,
        \be
        b(\sfrac{d}{dx})\left((e^x-1)\Delta(x)+C\right)=0.
        \label{eDE}
        \ee
        When $b(w)=(w-\lambda_1)^{K_1}
        \cdots(w-\lambda_\ell)^{K_\ell}$,
        the solutions are
                \be
        \Delta(x)={{\sum_{i=1}^\ell p_{K_i}(x)e^{\lambda_i x}-C}\over%
        {e^x-1}},~~~
        \deg p_{K_i}\leq K_i-1.
        \label{eDelta}
        \ee
\end{enumerate}
For the proof of these statements,
see reference \cite{rKR}.
%Explicit computation in terms of free fields
%can be found in \cite{rM}.

The independent components of the non-vanishing
\Winf generators may be taken as
$W(z^{-r}D^k)$ with $k=0,1,\ldots$, $\deg b_r-1$.
If the condition in the second item is satisfied,
the number of level $r$ non-vanishing generators is
$rK$ where $K\equiv$ $\deg b(w)$.
This observation immediately yields
the character of the Verma module,
\be
\ch(q,K)\equiv \mbox{tr }q^{L_0}=\chi(q)^K,
\qquad
\chi(q)=\prod_{j=1}^\infty
(1-q^j)^{-j}.
\label{eWIFC}
\ee
It already reveals one of  essential features
of the \Winf algebra.  A few years ago,
Cardy wrote an interesting paper on the modular invariance
in higher dimensions \cite{rC}.
By using $L_0=r \frac{\partial}{\partial r}$ as the
dilatation generator in the radial direction, he shows that
the partition function for the massless free fields in $d$
dimensions is written as,
\be
Z_d\equiv\mbox{tr }q^{L_0}=
\prod_{j=1}^\infty
(1-q^j)^{-D_d(j)},
\ee
where $D_d(j)$ is the number of spherical harmonics
with spin $j$
on $d-1$ sphere.        For large $j$, it behaves as
$D_d(j)\sim j^{d-2}$.
{}From this viewpoint, the character of the Verma module \refs{eWIFC}
has {\em three--dimensional} nature.
For this reason, its modular property is
quite different from those of Virasoro or Kac-Moody minimal
representations.
In particular, the character itself is not modular covariant.
However, if we write
\be
\log \chi(q=e^{-2\pi \delta})^K=\frac{K}{2\pi i}
\int_C \delta^{-s}F(s)ds,\quad
F(s)= (2\pi)^{-s}\Gamma(s)\zeta(s+1)\zeta(s-1),
\label{eZF}
\ee
and define a modified character,
\be
I(\delta)\equiv \frac{K}{2\pi i}
\int_C \delta^{-s}\frac{\Gamma(\frac{1}{2} s-\frac{1}{4})}{
\Gamma(\frac{1}{2} s+\frac{1}{4})}F(s+\sfrac{1}{2})ds,
\ee
then we can realize modular covariance even in three dimensions:
%to get modular covariance,
\be
I(\delta)=I(1/\delta)+RK(\delta^{-3/2}-\delta^{3/2}).
\ee
The second term on the right hand side may be
regarded as Casimir energy term.
%$R$ is some number.(??should be determined!!)
We remark that
it is not proportional to the central charge $C$
but is related to the degree $K$ of level one polynomial $b(w)$.

Later, we discuss the character formula
for the degenerate representation ($C=$integer).
In those cases, they are written by
free fields in two dimensions.
It illustrates the hybrid nature of
\Winf algebra.

%%%%%%%%%%%%%%%%%%%%%%%%%%%%%%%%%%%%%%%%
%                                      %
%  4. Determinant Formula              %
%                                      %
%%%%%%%%%%%%%%%%%%%%%%%%%%%%%%%%%%%%%%%%
\section{Determinant Formula}
After this preparation, we would like
to present our result on the Kac determinant for
quasi-finite representations.
We examine the simplest situations, {\em i.e.}, $K=1,2,3$.

When $K=1$, we write $b(w)=w-\lambda$, and thus \refs{eDelta} gives
$%\be
\Delta(x)=C{(e^{\lambda x}-1)}/{(e^x-1)}.
%\label{eDelta1}
%\ee
$
In this case, the highest weight is parametrized by
two constants $C,\lambda$.
For the first three levels, the relevant ket states
are,
\ba
\mbox{Level 1}&\quad&
W(z^{-1})|\lambda
\rangle\nonumber\\
\mbox{Level 2}&\quad&
W(z^{-2})|\lambda\rangle,\  W(z^{-1})^2|\lambda\rangle,\
W(z^{-2}D)|\lambda\rangle\nonumber\\
\mbox{Level 3}&\quad&
W(z^{-3})|\lambda\rangle,\ W(z^{-1})W(z^{-2})|\lambda\rangle,
W(z^{-1})^3|\lambda\rangle, \n
&\quad&
W(z^{-3}D)|\lambda\rangle,\
W(z^{-1})W(z^{-2}D)|\lambda\rangle,\
W(z^{-3}D^2)|\lambda\rangle.
\label{eBase}
\ea
Corresponding bra states may be given by changing
$z^{-r}$ into $z^r$.

In general, the number of relevant states grows as,
\ba
\chi(q)&=&1 + q + 3\,{q^2} + 6\,{q^3} + 13\,{q^4} + 24\,{q^5}
 + \nonumber\\
 && 48\,{q^6} + 86\,{q^7} + 160\,{q^8} + 282\,{q^9} +
  500\,{q^{10}} +\cdots
\ea
where the number of level $m$ non-vanishing states
is given by the coefficient of $q^m$.

We compute the Kac determinant up to level 8:
\ba
\det[1]&\propto& C\nonumber\\
\det[2]&\propto& C^3 (C-1)\nonumber\\
\det[3]&\propto& C^6 (C-1)^3(C-2)\nonumber\\
\det[4]&\propto& (C+1)C^{13} (C-1)^8(C-2)^3 (C-3)\nonumber\\
\det[5]&\propto& (C+1)^3C^{24} (C-1)^{17}(C-2)^8 (C-3)^3 (C-4)\nonumber\\
\det[6]&\propto& (C+1)^{10}C^{48}(C-1)^{37}(C-2)^{19} (C-3)^{8}
(C-4)^3(C-5)\nonumber\\
\det[7]&\propto& (C+1)^{23}C^{86}(C-1)^{71}(C-2)^{41} (C-3)^{19}
(C-4)^{8}(C-5)^{3}(C-6)\nonumber\\
\det[8]&\propto& (C+1)^{54}C^{161}(C-1)^{138}(C-2)^{85}
(C-3)^{43}\nonumber\\
&  &\times(C-4)^{19}(C-5)^{8}(C-6)^{3}(C-7).
\label{eDet1}
\ea
We remark that $\lambda$--dependent terms disappear
in the final expression due to nontrivial cancellations.
In the next section, we will explain why it happens
resorting to the spectral flow argument.
The maximal power of $C$ in $\det[m]$ is given by
the coefficient of $q^m$ in
$t\frac{d}{dt}\prod_{j=1}^{\infty}(1-tq^j)^{-j}|_{t=1}$.

When $K=2$, $b(w)=(w-\lambda_1)(w-\lambda_2)$ and we may parametrize
the solution of the differential equation as
\be
        \Delta(x)=C_1\frac{e^{\lambda_1x}-1}{e^x-1}+
        C_2\frac{e^{\lambda_2x}-1}{e^x-1},
        \qquad C=C_1+C_2.
\ee
The number of relevant states are generated by $\chi(q)^2$.
%\ba
%\chi(q)^2&=&1 + 2\,q + 7\,{q^2} + 18\,{q^3} + 47\,{q^4} + 110\,{q^5} +
%  258\,{q^6} +\nonumber\\
%&&   568\,{q^7} + 1237\,{q^8} + 2600\,{q^9} +
%  5380\,{q^{10}} + \cdots
%\ea
%
We calculate the determinant up to level $4$.
The final result is,
\ba
  \det[1]&\propto&(\lambda_1-\lambda_2)^2
  \prod_{k=1}^2 C_k \n
  \det[2]&\propto&(\lambda_1-\lambda_2)^{10}
  (\lambda_1-\lambda_2-1)^2 (\lambda_1-\lambda_2+1)^2
  \prod_{k=1}^2 C_k^4 (C_k-1) \n
  \det[3]&\propto&(\lambda_1-\lambda_2)^{34}
  \prod_{\epsilon=\pm 1}
  (\lambda_1-\lambda_2+\epsilon)^8(\lambda_1-\lambda_2+2\epsilon)^2
  \prod_{k=1}^2 C_k^{12}(C_k-1)^4(C_k-2) \n
%  (\lambda_1-\lambda_2-1)^{8} (\lambda_1-\lambda_2+1)^{8}\nonumber\\
% & &\times (\lambda_1-\lambda_2-2)^2(\lambda_1-\lambda_2+2)^2
%  F(C_1)F(C_2)\nonumber\\
%
  \det[4]&\propto&(\lambda_1-\lambda_2)^{108}
  \prod_{\epsilon=\pm 1}
  (\lambda_1-\lambda_2+\epsilon)^{30}(\lambda_1-\lambda_2+2\epsilon)^8
  (\lambda_1-\lambda_2+3\epsilon)^2 \n
  & & \times%\cdot
  \prod_{k=1}^2
 (C_k+1)C_k^{34}(C_k-1)^{14}(C_k-2)^4(C_k-3).
 \label{eDet2}
\ea
%Here the function $G(\lambda)$ is still to be determined.
In this case, $\lambda_1$ and $\lambda_2$ appear only through
their difference, $\lambda_1-\lambda_2$,
which will also be explained by spectral flow.
In the previous section, we noted that
something special happens when the difference of the roots
is integer. We can observe it explicitly by
the  appearance of zeros in the determinant.
%The appearance of zeros for it become integer
%comes from the reasoning we gave in the last section.

When two spins are identical, $\lambda_1=\lambda_2$, $\Delta(x)$ becomes
$(C(e^{\lambda x}-1)+p_1 x e^{\lambda x})/(e^x-1)$.
We may obtain the determinant formula for this case by writing,
%\be
$C_1=M+C$, %\quad
$C_2=-M$, %\quad
$\lambda_1=\lambda+\frac{p_1}{M}$, %\quad
$\lambda_2=\lambda$,
%\ee
and by taking a limit $M\rightarrow \infty$ \cite{rM}.
Although there are many other ways to reproduce the formula
for $\Delta(x)$, the Kac determinant itself does not depend on them,
\be
\det[1]\propto p_1^2,\quad
\det[2]\propto p_1^{10},\quad
\det[3]\propto p_1^{34},\quad
\det[4]\propto p_1^{108}.
\label{eDet2p}
\ee
We note that they have no dependence on $C$ or $\lambda$.

Finally when $K=3$, $b(w)=(w-\lambda_1)(w-\lambda_2)(w-\lambda_3)$,
the eigenvalues are given by
$
        \Delta(x)=\sum_{k=1}^3
        C_k{(e^{\lambda_kx}-1)}/{(e^x-1)},
        %C_2\frac{e^{\lambda_2x}-1}{e^x-1}+
        %C_3\frac{e^{\lambda_3x}-1}{e^x-1},
        %\qquad C=C_1+C_2+C_3.
        %\label{eDel3}
$
with $C=C_1+C_2+C_3$.
The number of relevant states is given by generating function
$\chi(q)^3$.
%\ba
%\chi(q)^3&=&1 + 3\,q + 12\,{q^2} + 37\,{q^3} + 111\,{q^4} + 303\,{q^5} +
%  804\,{q^6}\nonumber\\
%  &&+ 2022\,{q^7} + 4950\,{q^8} + 11715\,{q^9} +
%  27081\,{q^{10}} + \cdots.
%\ea
The Kac determinant for the first two levels are,
\ba
\det[1]&\propto&\prod_{i<j}(\lambda_i-\lambda_j)^2
 \prod_{k=1}^3 C_k\label{eDet3}\\
\det[2]&\propto& \prod_{i<j}(\lambda_i-\lambda_j)^{12}
(\lambda_i-\lambda_j-1)^2(\lambda_i-\lambda_j+1)^2
        \prod_{k=1}^3 C_k^5 (C_k-1).\nonumber
%\label{eDet3}
\ea

When the $\lambda_i$'s are identical, the generating function
becomes,
\be
\Delta(x)=\frac{(C+p_1 x+p_2 x^2) e^{\lambda x}-C}{e^x-1}.
\label{eDelta3}
\ee
This solution can be derived  by
putting
$C_1=M^2$, $C_2=-2M^2+C$, $C_3=M^2$,
$\lambda_2=\lambda$, $\lambda_1=\lambda+\frac{\sqrt{p_2}}{M}
+\frac{1}{2M^2}p_1$,
$\lambda_3=\lambda-\frac{\sqrt{p_2}}{M}+\frac{1}{2M^2}p_1$,
and taking a limit $M\rightarrow\infty$.  Again, there are many
other ways to get the same limit,
while the Kac determinant itself does not depend on them,
\be
\det[1]\propto p_2^3,\qquad
\det[2]\propto p_2^{18}.
\ee
Not that they are free of other parameters $(\lambda,C,p_1)$.

By inspecting these formulae,
one may summarize the properties of the Kac determinant for
the quasi-finite representation of \Winf algebra as follows:
\begin{enumerate}
        \item  The determinant is always factorized into two parts.
        The first depends only on $C$ and the second on
        $\lambda_i-\lambda_j$.

        \item  The additional null states appear only when
        $C$ is integer or $\lambda_i-\lambda_j$ is integer.

        \item  When some $\lambda_i$'s are identical, only the top
        component of $p_{K}(x)$ in \refs{eDelta} is relevant
        to the Kac determinant. In particular, when it does not
        vanish, there are no additional null states.
\end{enumerate}
%We conjecture that these features
%hold true in all orders.

%%%%%%%%%%%%%%%%%%%%%%%%%%%%%%%%%%%%%%%%
%                                      %
%  5. Spectral Flow                    %
%                                      %
%%%%%%%%%%%%%%%%%%%%%%%%%%%%%%%%%%%%%%%%

\section{Spectral Flow}
In the previous section,
we see that the Kac determinants depend only
on the difference between the parameters $\lambda_i$.
This fact is naturally understood if we notice the existence of
one parameter family of automorphisms (spectral flow) in \Winf algebra
\cite{rPRS}.
%Other representations, $\lambda\neq 0$,
%are obtained by using the spectral flow,
%an automorphism of $W_{1+\infty}$.

The transformation rule is given by
\be
  W^\lambda(z^re^{xD})=W(z^r e^{x(D+\lambda)})
  +C\frac{e^{\lambda x}-1}{1-e^x}\delta_{n0},
\ee
where $\lambda$ is an arbitrary parameter.
For lower components, for example, it is expressed as
\ba
  J'_n&\!\!=\!\!&J_n-\lambda C\delta_{n0},\n
  L'_n&\!\!=\!\!&L_n-\lambda J_n
  +\sfrac{1}{2}\lambda(\lambda-1)C\delta_{n0}.
\ea
One can easily show that $W^{\lambda}(\cdot)$
satisfies the same commutation relation with \refs{comm}.
Furthermore, a highest weight state $|\Lambda\rangle$ with respect to
the original generators $W(z^r D^k)$ is also a highest weight state
with respect to the new generators $W^\lambda(z^r D^k)$, while the
minimal polynomial $b(w)$ and the weight $\Delta(x)$ are replaced,
respectively, by $b(w-\lambda)$ and
\be
  e^{\lambda x}\Delta(x)+C\frac{e^{\lambda x}-1}{e^x-1}.
\ee
%We derive that $W^{\lambda}(\cdot)$
%satisfy the same commutation
%relation \refs{comm}.
%Let
%$|\Lambda\rangle$
%be the highest weight state with
%respect to generators $W(z^r D^k)$ with the minimal
%polynomial $b(w)$ and the weight $\Delta(x)$.
%Then we show that $|\Lambda\rangle$
%is also the highest weight representation
%with respect to generators $W^\lambda(z^r D^k)$,
%with the minimal polynomial $b(w-\lambda)$ and the weight
%\be
%  e^{\lambda x}\Delta(x)+C\frac{e^{\lambda x}-1}{e^x-1}.
%\ee
This implies that the spectral flow transforms $\lambda_i$ into
$\lambda_i+\lambda$.
On the other hand, any automorphism does not change determinants.
Thus, we conclude that the Kac determinant depends on the $\lambda_i's$
only through their differences.

%%%%%%%%%%%%%%%%%%%%%%%%%%%%%%%%%%%%%%%%
%                                      %
%  6. Characters for Degenerate        %
%     Representations                  %
%                                      %
%%%%%%%%%%%%%%%%%%%%%%%%%%%%%%%%%%%%%%%%
\section{Characters for Degenerate Representations}

{}From the Kac determinant, we can extract
the information on null states in the Verma module
with integer central charges.
{}From the preceding discussions, it is plausible to assume
%In general, we conjecture
the following form of Kac determinant for $K=1$, $b(w)=w-\lambda$:
\be
        \det[N]=\prod_{n=-\infty}^{\infty} (C-n)^{\tilde\phi(n,N)},\qquad
        \tilde\phi(n,N)\ge\phi(n,N),
        \label{eDetG}
\ee
where $\phi(n,N)$ is the number of null states at level $N$
in the $C=n$ highest weight representation.
Comparing the character of the Verma module,
one may derive,\footnote{
For $C=0$ case, $\ch_0(q)=1$.}
\be
  \ch_n(q)\equiv\mbox{tr } q^{L_0}=
  q^{\frac{1}{2}n\lambda(\lambda-1)}
  \Bigl(
  \chi(q)-\sum_{N=0}^\infty \phi(n,N)q^N
  \Bigr).
\ee

We now would like to propose the character formulae for $C=n\geq -1$
degenerate representation with $b(w)=w-\lambda$,
which is consistent with the Kac determinant above:
\ba
  \ch_{n}(q)&\!\!=\!\!&
  q^{\frac{1}{2}n\lambda(\lambda-1)}
  \prod_{j=1}^{\infty}\frac{1}{(1-q^j)(1-q^{j+1})\cdots(1-q^{j+n-1})}
  \qquad (C=n>0),
  \label{eChF}\\
  \ch_{-1}(q)&\!\!=\!\!&
  q^{-\frac{1}{2}\lambda(\lambda-1)}
  \prod_{j=1}^{\infty}\frac{1}{(1-q^j)^2}\cdot
  \sum_{m=0}^{\infty}(-1)^mq^{\frac{1}{2}m(m+1)} \qquad
  (C=-1).
  \label{eChB}
\ea
Recall that the former describes the $n$ free--fermion system,
while the latter the bosonic ghost.
Actually, in these cases we can construct the full character formula
as will be shown later, and \refs{eChB} is obtained from the full
character by restricting it such as counting the conformal weight only.

We here mention a remarkable similarity
between the $C=1$ and $C=-1$ characters.
In fact, a short calculation shows the following equations:
\ba
  \ch_1(q)&\!\!=\!\!&
  q^{\frac{1}{2}\lambda(\lambda-1)}
  \sum_{m=0}^{\infty}q^m
  \Bigl(q^{\frac{1}{2}m(m-1)}\prod_{j=1}^m\frac{1}{1-q^j}\Bigr)^2, \\
  \ch_{-1}(q)&\!\!=\!\!&
  q^{-\frac{1}{2}\lambda(\lambda-1)}
  \sum_{m=0}^{\infty}q^m
  \Bigl(\prod_{j=1}^m\frac{1}{1-q^j}\Bigr)^2.
\ea
They may suggest that there exists a general correspondence between
bosonic and fermionic characters.

The character formula \refs{eChF} is a simple consequence of
the following observation:
{\em  The representation space of \Winf algebra with
$C=n>0$, is generated by $W(z^{-r}D^k)$, with
$0\leq k \leq n-1$, $r\geq k+1$.}
We confirmed this statement for all generators in the case of
$C=1$, and for lower generators in the case of $C=2$.
Although we have no rigorous proof at this stage, \refs{eChF}
is consistent with the Kac determinant obtained above.
In the appendix, we describe the general structure
of null states for the free field theories.
%It is simple and should be useful in the future.

The character formula \refs{eChB} for $C=-1$ is rigorously proved as follows.
Since we have already demonstrated the spectral flow,
we can restrict ourselves to the simplest case, $b(w)=w$.
We follow the argument of \cite{rAFOQ}:
The basis of the representation space are
$\beta_{-r_1}\cdots\beta_{-r_k}
\gamma_{-s_1}\cdots\gamma_{-s_k}|0\rangle$
with $r_1\geq\cdots\geq r_k\geq 1$ and $s_1\geq\cdots\geq s_k\geq 0$.
These states are simultaneous eigenstates of $W(D^k)$, since
\be
  \lbrack W(D^k),\beta_r\rbrack=\bar{a}^k_r\beta_r,~~~
  \lbrack W(D^k),\gamma_s\rbrack=a^k_s\gamma_s,
\ee
%where $\bar{a}^k_r=(r-\lambda)^k$ and $a^k_s=-(-s-\lambda)^k$.
where $\bar{a}^k_r=r^k$ and $a^k_s=-(-s)^k$.
Thus, the full character is calculated as
\be
  \ch_{-1}^{\lambda=0}\equiv
  {\rm tr}\prod_{k=0}^{\infty}x_k^{W(D^k)}
  = t^0 \mbox{~term of~}
%  \prod_{k=1}^{\infty}x_k^{\Delta_k}
  \prod_{r=1}^{\infty}
  \Bigl(1-t\prod_{k=0}^{\infty}x_k^{\bar{a}^k_{-r}}\Bigr)^{-1}
  \prod_{s=0}^{\infty}
  \Bigl(1-t^{-1}\prod_{k=0}^{\infty}x_k^{a^k_{-s}}\Bigr)^{-1},
\ee
from which the character \refs{eChB} is obtained as
\ba
  \ch_{-1}^{\lambda=0}(q)&\!\!=\!\!&
  t^0 \mbox{~term of~}
%  q^{\frac{1}{2}\lambda(\lambda-1)}
  \prod_{r=1}^{\infty}(1-tq^r)^{-1}
  \prod_{s=0}^{\infty}(1-t^{-1}q^s)^{-1} \n
  &\!\!=\!\!&
%  q^{\frac{1}{2}\lambda(\lambda-1)}
  \sum_{\ell=0}^{\infty}\ch^{W^{c=2}_{\infty}}_{\ell}
  =
%  q^{\frac{1}{2}\lambda(\lambda-1)}
  \prod_{j=1}^{\infty}\frac{1}{(1-q^j)^2}\cdot
  \sum_{m=0}^{\infty}(-1)^mq^{\frac{1}{2}m(m+1)}. \nonumber
\ea
Here we have used the character of $W_{\infty}$ algebra
with $c=2$ \cite{rO}.

%%%%%%%%%%%%%%%%%%%%%%%%%%%%%%%%%%%%%%%%
%                                      %
%  7. Future Issues                    %
%                                      %
%%%%%%%%%%%%%%%%%%%%%%%%%%%%%%%%%%%%%%%%
\section{Future Issues}
In this letter,
we examined the Kac determinant of \Winf algebra for the first few levels.
Although the information we obtained is still restricted,
it helps us to understand the structure of
degenerate representations and their characters.
In particular, our result seems to suggest
that non-trivial representation happens only when $C$ is an integer.

There are many things which are worth thorough investigation.
As mathematical problems,
besides the completion of rigorous proofs for the above formulae
on the Kac determinants and characters,
it must be interesting to investigate the system
where $\lambda_i-\lambda_j$ is integer.
As problems of physics,
we should clarify the role of \Winf algebra in three--dimensional systems,
and the origin of the disappearance of the modular invariance
in ordinary sense.
We hope to report on the above subjects in our future issues.

\vskip 5mm
\noindent{\bf Acknowledgements:}
Two of the authors (Y.M. and S.O.) would like
to thank members of YITP for their hospitality
where part of this work was carried out.
Y.M. and S.O. are obliged to Prof.\ Kenzo Inoue and Prof.\ Takeo Inami
for financial support.
This work is also supported in part by
Grant-in-Aid for Scientific Research from Ministry of Science and Culture,
and by Soryushi-Shogakkai.

%%%%%%%%%%%%%%%%%%%%%%%%%%%%%%%%%%%%%%%%
%                                      %
%  Apdx:  Null States for Free Fields  %
%                                      %
%%%%%%%%%%%%%%%%%%%%%%%%%%%%%%%%%%%%%%%%
\vskip 5mm

\noindent{\bf Appendix: Null States for Free Fields}

\noindent In this appendix,
we show a general characterization of null states of free field theories.
We treat bosonic and fermionic cases in parallel fashion,
{\em i.e.}, ${\bf b}(z)=\sum_{r\in{\bf Z}}{\bf b}_rz^{-r-1+\lambda}$
stands for $\beta$ or $b$, and
${\bf c}(z)=\sum_{r\in{\bf Z}}{\bf c}_rz^{-r-\lambda}$
stands for $\gamma$ or $c$, depending on which we treat,
free boson ($\epsilon=-1$) or fermion ($\epsilon=1$).
%We also introduce a parameter $\epsilon$ which takes $+1$
%when we are discussing bosonic case and $-1$ for the fermionic case.
$|\lambda\rangle$ is characterized by
${\bf b}_r|\lambda\rangle=$${\bf c}_s|\lambda\rangle=0$
($r\geq 0$, $s\geq 1$).
The \Winf algebra with central charge
$C=\epsilon n$ is realized by
$W(z^rf(D))$$=$$\sum_{\alpha=1}^n \oint\frac{dz}{2\pi i}
:{\bf b}^{(\alpha)}(z)z^rf(D){\bf c}^{(\alpha)}(z):$.
In the representation space,
the element of \Winf module is generated by
$
  E(r,s)=\sum_{\alpha=1}^n
  {\bf b}^{(\alpha)}_{-r}{\bf c}^{(\alpha)}_{-s},
%  \label{eErs}
$
where $r\geq 1$ and $s\geq 0$.

We claim that the following operator for $C=\epsilon n$ becomes null,
\be
  \mbox{$\det_{\epsilon}$}
  \left(
  \begin{array}{ccc}
    E(r_1,s_1)&\cdots&E(r_1,s_{n+1})\\
    \vdots&&\vdots\\
    E(r_{n+1},s_1)&\cdots&E(r_{n+1},s_{n+1})
  \end{array}
  \right),
  \label{null}
\ee
where $r_1\leq\cdots\leq r_{n+1}$, $s_1\leq\cdots\leq s_{n+1}$
for fermionic cases, and
$r_1<\cdots<r_{n+1}$, $s_1<\cdots<s_{n+1}$ for bosonic cases.
$\det_{\epsilon}A$ stands for a permanent\footnote{
Permanent of a $m\times m$ matrix $A$ is defined by
$\det_{+}(A)\equiv \sum_{\sigma}\prod_i A_{i\sigma(i)}$,
where $\sigma$ is the permutations of the set $\{1,2,\cdots,m\}$.}
(determinant) of a matrix $A$ for
$\epsilon=+1 ~(\epsilon=-1)$, respectively.

The proof of this statement is straightforward.
By definition, the  $m\times m$ matrix $(E(r_i,s_j))$ is a product of
the $m\times n$ matrix $({\bf b}^{(\alpha)}_{r_i})$ and
the $n\times m$ matrix $({\bf c}^{(\alpha)}_{s_j})$.
If $m>n$, $\det_{\pm}$ vanishes since
the matrix $(E(r_i,s_j))$ becomes a projection operator.
%For a $n\times m$ matrix $A$ and a $m\times n$ matrix $B$,
%$\det_{\pm}(AB)$ vanishes if $A$ and $B$ have Grassmann odd(even)
%elements and $n<m$. Therefore \refs{null} is a null operator.

{}From this observation, the appearance of first null state for the
fermionic case should occur at level $n+1$ for $C=n$ and
at level $(n+1)^2$ for the bosonic case $C=-n$.
We can easily confirm it in our determinant formula.

%%%%%%%%%%%%%%%%%%%%%%%%%%%%%%%%%%%%%%%%
%                                      %
%   References                         %
%                                      %
%%%%%%%%%%%%%%%%%%%%%%%%%%%%%%%%%%%%%%%%

\end{document}